\documentclass[%
 reprint,
 amsmath,amssymb,
 aps,
]{revtex4-1}

\usepackage{graphicx}% Include figure files
\usepackage{dcolumn}% Align table columns on decimal point
\usepackage{bm}% bold math
\usepackage{subfigure}
\begin{document}
\preprint{APS/123-QED}

\title{Light Source Monitoring in Quantum Key Distribution with Single Photon Detector at Room Temperature}% Force line breaks with \\

\author{Gan Wang$^1$}
\author{Zhengyu Li$^1$}
\author{Yucheng Qiao$^1$}
\author{Ziyang Chen$^1$}
\author{Xiang Peng$^1$}
\author{Hong Guo$^1$}

\thanks{Corresponding author: hongguo@pku.edu.cn.} %\\$^{\dag}$Corresponding author: xiangpeng@pku.edu.cn.}

\affiliation{$^1$State Key Laboratory of Advanced Optical Communication Systems and Networks, School of Electronics Engineering and Computer Science, and Center for Quantum Information Technology, Peking University, Beijing 100871, China}
%\affiliation{$^2$Science and Technology on Security Communication Laboratory, Institute of Southwestern Communication, Chengdu 610041, China}

\begin{abstract}

Photon number resolving monitoring is a practical light source monitoring scheme in QKD systems, which reduces the impacts from untrusted sources effectively. This scheme requires a single photon detector, normally working at low temperature to suppress its dark count rate. In this paper, we use a room-temperature detector and show that the dark count rate is irrelevant to the monitoring performance in our scheme, which can sufficiently relax requirements on the detector's working conditions as well as integration complexity, and this would be highly demanded for practical systems. Furthermore, influences of parameter drifts at room temperature are analyzed, and the monitoring scheme is testified in a real QKD system.

\end{abstract}

%\pacs{Valid PACS appear here}% PACS, the Physics and Astronomy
                             % Classification Scheme.
%\keywords{Suggested keywords}%Use showkeys class option if keyword
                              %display desired
\maketitle

%\tableofcontents

\section{Introduction}
Quantum key distribution (QKD) is a theoretically secure method to distribute secret keys between the information transmitter (Alice) and the receiver (Bob). After more than three decades' development, QKD is coming to the edge of commercial applications. However, in real QKD systems, loopholes in optical components may cause practical security issues, among which the loopholes in quantum sources and detectors are the most vulnerable ones~\cite{lutkenhaus2000security,vakhitov2001large,gisin2006trojan,xu2010experimental,makarov2006effects,zhao2008quantum,lydersen2010hacking,gerhardt2010full}. Recently-proposed measurement-device-independent protocol has removed the loopholes in detectors with practical techniques~\cite{lo2012measurement,da2013proof,rubenok2013real,liu2013experimental,tang2014experimental}, while those in sources still remain. Normally, weak coherent sources, rather than ideal single photon sources, are used in practical systems, and decoy state protocol is adopted to reduce multi-photon imperfections~\cite{lo2005decoy,wang2005beating}. Under these conditions, the practical security of QKD systems is guaranteed by trusted sources, whose photon number distribution is known to Alice. In practice, trusted sources are not always available especially under the attack of eavesdroppers, thus light source monitoring (LSM) is indispensable~\cite{wang2007simple,peng2008experimental,wang2008general,zhao2008quantum2}.

Commonly, there are three kinds of LSM methods, including average photon number monitoring~\cite{gisin2006trojan,peng2008experimental}, `Untagged bits' monitoring~\cite{zhao2008quantum2,peng2010passive,zhao2010security}, and photon number resolving (PNR) monitoring~\cite{xu2010passive,xu2012sarg}, among which PNR is the most precise. Ideal PNR monitoring requires a photon number resolving detector, which can be replaced by a single photon detector (SPD) with a variable attenuator in practical scheme. The SPD used in a QKD system is normally an InGaAs avalanche photon diode detector, being cooled down by a thermo-electric cooler and working below $-30^{\circ}$C$\sim$$-50^{\circ}$C to suppress its dark count rate (DCR)~\cite{eraerds2010quantum,dixon2010continuous,lucamarini2013efficient,comandar2014room}. However, when working at room temperature, an InGaAs SPD will have a high DCR~\cite{comandar2014room,pellegrini2006design}, which is not suitable for the receiver of QKD system but has potential of application in LSM.

In this paper, we propose and realize PNR LSM based on a room-temperature SPD. This scheme reduces the requirements of the SPD's working conditions as well as its size and power consumptions significantly by removing cooler from the module, which makes it much easier to integrate the monitoring module in a QKD system. We prove that high DCR at room temperature leads to no degradation of the performance of monitoring, thus guarantees high security for practical QKD systems. Influences of parameter drifts as well as modifications on the mathematical model are also presented. Furthermore, our scheme is testified with a real QKD system.

This paper is organized as follows: In Sec.~\ref{Sec2-1}, we discuss the mechanisms of why the performance of our LSM scheme is immune to high DCR at room temperature. In Sec.~\ref{Sec2-2}, effects of parameter drifts and the corresponding modifications on the mathematical model are presented. In Sec.~\ref{Sec3}, we measure the ranges of the parameter drifts in real detectors and realize PNR LSM with a room temperature SPD in a practical QKD system.

\section{LSM with a PNR Detector at Room Temperature}\label{Sec2}

In this section, we start from a review of the basic theory of PNR LSM and then extend the discussion to the case where a room-temperature SPD is used, and also analyze the potential influences due to the parameter drifts of SPD.

\subsection{LSM with a PNR Detector}\label{Sec2-1}

In BB84 protocol, according to the GLLP theory, the secure key rate $R$ of a QKD system is~\cite{gllp2004}
\begin{equation}
  R = \frac{1}{2}Q\{ \Delta_1[1-H_2(e_1)] - H_2(E) \},
\end{equation}
where $Q$ and $E$ indicate the count rate and quantum bit error rate (QBER) in a QKD system, $\Delta_1$ and $e_1$ indicate the probability and bit error rate of single-photon pulses.

Decoy state protocol was introduced to estimate $\Delta_1$ and $e_1$ more precisely, offering an effective way to deal with the multi-photon problem caused by weak coherent sources~\cite{lo2005decoy,wang2005beating,ma2005practical}. In double decoy states protocol, the transmitter Alice modulates the intensity of light pulses randomly into three states: signal, decoy, and vacuum state, whose light fields are
\begin{equation}
\begin{aligned}
&   \rho_{\rm{signal}}=\sum_{n=0}^{\infty}a_n|n\rangle\langle n|, \\   %\tag{1.1}
&   \rho_{\rm{decoy}}=\sum_{n=0}^{\infty}a'_n|n\rangle\langle n|, \\
&   \rho_{\rm{vacuum}}=|0\rangle\langle 0|,
\end{aligned}
\end{equation}
where $a_n$ and $a'_n$ are the coefficients of Fock state $|n\rangle$, assumed to be known and controlled by Alice. However, in practical QKD systems, the untrusted source problem still remains, and a lower bound of $\Delta_1$ is given by~\cite{wang2008general}
\begin{equation}
  \Delta_1 \geq \frac{a^L_1 (a^L_2 Q_d - a'^U_2 Q_s - a^L_2 a'^U_0 Y_0 + a'^U_2 a^L_0 Y_0)}{Q_s(a'^U_1 a^L_2 - a^L_1 a^U_2)},
\end{equation}
where $a^U_n$ ($a^L_n$) is the upper (lower) bound of $n$ photons' probability in signal state and $a'^U_n$ ($a'^L_n$) is the corresponding bound in decoy state, which needs to be monitored in real regime by LSM.

\begin{figure}[h]
  \centering
  \includegraphics[width=9cm]{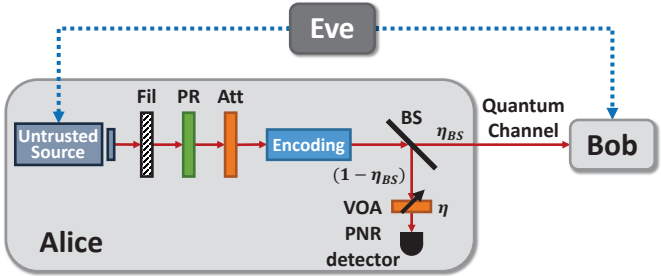}
  \caption{Practical passive PNR LSM scheme. The untrusted source in the information transmitter (Alice) may be influenced or controlled by the eavesdropper (Eve). After being emitted from the source, a light pulse passes through an optical filter (Fil), a phase randomizer (PR), and an attenuator (Att). The pulse will be encoded and a beam splitter (BS) with a transmittance $\eta_{BS}$ separates it into two beams. One goes through a variable optical attenuator (VOA) with a transmittance $\eta$ to the monitoring SPD, and the other is sent through the quantum channel to the receiver (Bob).}\label{1-PNR}
\end{figure}

The ideal PNR LSM theory requires a photon number resolving detector, but a practical SPD is used in real situations to estimate $\{ a^U_n, a^L_n, a'^U_n, a'^L_n\}$. Among the practical PNR LSM schemes, the passive scheme is a more practical one, as shown in Fig. \ref{1-PNR}~\cite{xu2010passive}.

\begin{figure}[t]
  \centering
  \includegraphics[width=8cm]{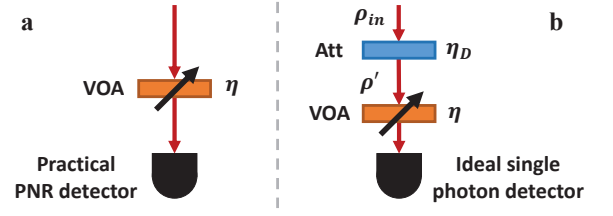}
  \caption{Practical establishment of PNR monitoring (a) is equivalent to the ideal model (b). A practical PNR detector can be regarded as an `ideal' SPD with an attenuator ($\eta_D$) in front of it.}\label{1-model}
\end{figure}

In this scheme, the practical SPD is equivalent to an `ideal' one with an attenuator with transmittance $\eta_D$ in front of it (Fig. \ref{1-model}). An `Ideal' detector here is defined to have a quantum detection efficiency of 100\% with nonzero DCR.
$\eta_{BS}$ is the transmittance of BS and we set
\begin{equation}
  \eta_{BS} = (1-\eta_{BS}) \eta_D,
\end{equation}
so that the photon number distribution of the light field before entering the quantum channel and that of $\rho'$ in Fig. \ref{1-model} are the same. Thus, the light field $\rho'$ in signal state reads
\begin{equation}
  \rho'_{\rm{signal}} = \sum_{n=0}^{\infty}a_n|n\rangle\langle n|,
\end{equation}
while $\rho'$ in decoy state reads
\begin{equation}
  \rho'_{\rm{decoy}} = \sum_{n=0}^{\infty}a'_n|n\rangle\langle n|.
\end{equation}

In signal state, when the light beam passes the VOA with transmittance of $\eta$, the probability of detecting zero photon with the detector is
\begin{equation}
  P(\eta) = (1-\lambda) \sum_{n=0}^{\infty} (1-\eta)^n a_n = (1-\lambda) P_0(\eta),
\end{equation}
where $\lambda$ is the DCR of SPD. $P_0(\eta)$ is the probability of detecting zero photon with the `ideal' detector, being irrelevant to $\lambda$.

If Alice chooses different $\eta\in\{\eta_0, \eta_1, \eta_2\}~(\eta_0>\eta_1>\eta_2)$ randomly, $\{ a^U_n, a^L_n, a'^U_n, a'^L_n\}$ can be calculated with $P(\eta)$ under different situations, and $P(\eta)$ is measured with the count rates of the monitoring detector. We set $\eta_0=1$, thus $\{a^U_n, a^L_n\}$ are given by (more detailed calculations are shown in Appendix A.)
\begin{equation}
\begin{aligned}
&  P_0(\eta_0)=\frac{P(\eta_0)}{1-\lambda}, \\
&  a^U_0=a^L_0=P_0(\eta_0),\\
&  a^U_1=\frac{P_0(\eta_1)-P_0(\eta_0)}{1-\eta_1},\\
&  a^L_1=\frac{P_0(\eta_1)-P_0(\eta_0)[1-(1-\eta_1)^2]-(1-\eta_1)^2}{(1-\eta_1)-(1-\eta_1)^2},\\
&  a^U_2=\frac{P_0(\eta_2)-P_0(\eta_0)-(1-\eta_2)a^L_1}{(1-\eta_2)^2},\\
&  a^L_2=\frac{P_0(\eta_2)-P_0(\eta_0)[1-(1-\eta_2)^3]}{(1-\eta_2)^2-(1-\eta_2)^3}\\
&  -\frac{[1-\eta_2-(1-\eta_2)^3]a^U_1-(1-\eta_2)^3}{(1-\eta_2)^2-(1-\eta_2)^3}.
\end{aligned}
\label{eq1}
\end{equation}
Similarly, we can get $\{a'^U_n, a'^L_n\}$ in decoy state. The above equations prove that when $\lambda$ is known and stable to Alice, $\{a^U_n, a^L_n\}$ and $\{a'^U_n, a'^L_n\}$ can be computed even though $\lambda$ is much higher than its value in low-temperature regime.

Typically, an InGaAs avalanche photon diode detector works under $-30^{\circ}$C$\sim$$-50^{\circ}$C to suppress its DCR~\cite{eraerds2010quantum,dixon2010continuous,lucamarini2013efficient,comandar2014room}. When working at 1550 nm and having a detection efficiency of 10\%, the DCR of the SPD is $10^{-7}\sim10^{-5}$ per pulse at low temperature, but it will rise to a relatively high level at room temperature~\cite{peng2007experimental,yuan2009practical,takemoto2010transmission,sasaki2011field,wang2014field,comandar2014room}. In our experiment, the SPD has a DCR around $5.4\times 10^{-6}$ per pulse at low temperature, while $5.8\times 10^{-4}$ at room temperature. A high DCR will increase the time of calculating monitoring results, but as $\lambda \ll 1$ still holds at room temperature, the increase of time is negligible. Meanwhile, working at room temperature not only reduces the requirements of working conditions of the SPD, but also decreases its size and power consumptions by simplifying (even removing) its temperature control, thus makes the monitoring detector much easier to be integrated in QKD systems as an important complement enhancing its practical security.

In the following part, modifications of $\{a^{U(L)}_n, a'^{(U)L}_n\}$ are discussed when considering the parameter drifts of practical SPD. Accoring to current QKD experiments~\cite{Zhao2006,Dixon08,Yin2008,Liu10}, in the simulation and experiment below, we set $\mu=0.6,~\nu=0.1$. $\eta_1,\eta_2$ are chosen from these pairs of values: $\{\eta_1=0.9, \eta_2=0.1; \eta_1=0.9, \eta_2=0.5; \eta_1=0.8, \eta_2=0.1; \eta_1=0.8, \eta_2=0.2\}$ (the discussion of the values of $\eta_1$ and $\eta_2$ can be found in Appendix A).

\subsection{Fluctuation of Dark Count Rate and Detection Efficiency}\label{Sec2-2}

In our LSM scheme, the DCR of the room-temperature SPD is required to be known and stable, but it inevitably fluctuates in actual environments. Other parameters, especially the detection efficiency of SPD, may also drift. If the parameters drifts are ignored, the final secret key rate could be overestimated. To solve this problem, one method is to monitor all the parameters of the detector in real time and calculate secret key rate with actual values. This will increase the system complexity obviously, and moreover, it is difficult to monitor all the parameters continuously. A more simple solution is to measure the range of parameter drifts and to modify $\{a^U_n, a^L_n, a'^U_n, a'^L_n\}$ with the worst cases within the fluctuation range, which gives a reasonable estimation of the lower bound of secret key rate $R$. In the following part, the derivation of modified mathematical model with parameter drifts as well as their valid range are presented. The modification results when using a real room-temperature SPD will be discussed in the experiment section.

Considering the drift of the DCR of SPD, we denote the original DCR as $\lambda$, while $\lambda'$ is the actual DCR with drifting, and $\lambda' = (1+\delta) \lambda$, where $|\delta\lambda| \ll 1$. $\{a^U_n, a^L_n\}$ are the bounds calculated with $\lambda$ and $\{a^U_{n,DCR}, a^L_{n,DCR}\}$ are the bounds with $\lambda'$. $\{a^U_{n,DCR}, a^L_{n,DCR}\}~(n=0,1,2)$ are given by (more detailed derivations are shown in Appendix B)
\begin{equation}
\begin{aligned}
&  a^U_{0,DCR}=a^L_{0,DCR}\approx a^U_0 (1+\delta\lambda),\\
&  a^U_{1,DCR}\approx a^U_1 (1+\delta\lambda),\\
&  a^L_{1,DCR}\approx a^L_1 (1+\delta\lambda)-\delta\lambda\frac{1-\eta_1}{\eta_1},\\
&  a^U_{2,DCR}\approx a^U_2 (1+\delta\lambda),\\
&  a^L_{2,DCR}\approx a^L_2 (1+\delta\lambda)-\delta\lambda\frac{1-\eta_2}{\eta_2}.
\end{aligned}
\label{eq2}
\end{equation}

\begin{figure}[h]
  \centering
  \includegraphics[width=8cm]{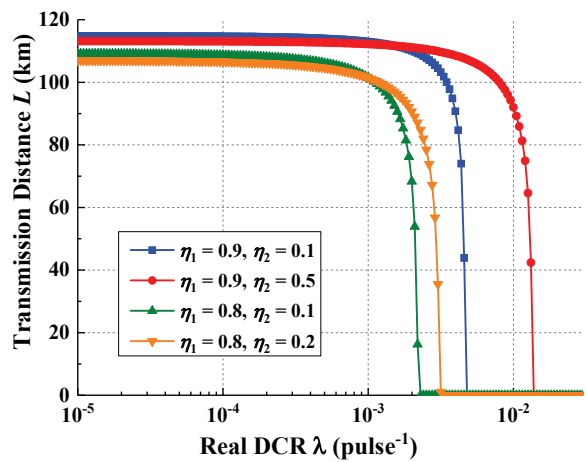}
  \caption{Transmission distance $L$ of a QKD system when the monitoring SPD's DCR $\lambda$ drifts. The original DCR $\lambda=5\times 10^{-4}$ at room temperature.}\label{2-DCR}
\end{figure}

Using Eq. \ref{eq2}, we calculate the system's secret key rate $R$ and the maximum transmission distance $L$ when DCR drifts, as in Fig. \ref{2-DCR}~\cite{footnote1}.

The original DCR $\lambda$ of room-temperature SPD is $5\times 10^{-4}$ per pulse, close to the DCR in actual environment. $L$ remains relatively stable when $\lambda$ decreases, while decreases when $\lambda$ increases and falls to zero when $\lambda$ is high enough. However, even when $\lambda = 10^{-3}$, $L$ only decreases by 1 km as for $\{\eta_1=0.9, \eta_2=0.1\}$, and decreases by approximately 10 km when $\lambda$ rises to $3\times 10^{-3}$. As for the worst case $\{\eta_1=0.8, \eta_2=0.1\}$, $L$ decreases by 6 km when $\lambda = 10^{-3}$, while descends rapidly with the further increase of $\lambda$ and the system cannot generate keys when $\lambda > 2.2\times 10^{-3}$. In practice, the DCR of the SPD with a detection efficiency 10\% never reaches $10^{-3}$ at room temperature, so the decline of $L$ is within 10 km, an acceptable range in practical implementation. We also observe that when $\eta_1$ is fixed, the larger $\eta_2$ is, the larger tolerance the system has for the drift of DCR.

When the detection efficiency of the SPD varies with time, the modification method should be discussed under different conditions of both the increasing and the decreasing efficiency. Let $\eta_D$ represent the original detection efficiency and $\eta'_D$ be the actual efficiency with varying. When $\eta_D$ rises to $\eta'_D$, we have $\eta'_D=(1+\delta)\eta_D$, where $|\delta\eta_D| \ll 1$. $\{a^U_n, a^L_n\}$ still represent the bounds calculated with $\eta_D$, and $\{a^U_{n,\eta_D\uparrow}, a^L_{n,\eta_D\uparrow}\}$ are the actual bounds when $\eta_D$ rises to $\eta'_D$. Thus, $\{a^U_{n,\eta_D\uparrow}, a^L_{n,\eta_D\uparrow}\}$ read as
\begin{equation}
\begin{aligned}
 &  a^U_{0,\eta_D\uparrow}\approx a^U_0+a^U_1\delta+a^U_2\cdot\frac{\delta^2}{1-\delta},  \\
 &  a^L_{0,\eta_D\uparrow}\approx a^L_0+a^L_1\delta+a^L_2\cdot\delta^2,  \\
 &  a^U_{1,\eta_D\uparrow}\approx (1-\delta)[a^U_1+a^U_2\cdot\frac{2\delta-\delta^2}{(1-\delta)^2}],  \\
 &  a^L_{1,\eta_D\uparrow}\approx (1-\delta)[a^L_1+a^L_2\cdot2\delta],  \\
 &  a^U_{2,\eta_D\uparrow}\approx \frac{a^U_2}{1-\delta},  \\
 &  a^L_{2,\eta_D\uparrow}\approx a^L_2\cdot(1-\delta)^2.
\end{aligned}
\end{equation}

When $\eta_D$ falls to $\eta'_D$, we have $\eta_D=(1-\sigma)\eta'_D$, which is equivalent to $\eta_D=(1+\delta)\eta'_D$, where $1+\delta=\frac{1}{1-\sigma}, |\delta\eta'_D| \ll 1$. Then, $\{a^U_{n,\eta_D\downarrow}, a^L_{n,\eta_D\downarrow}\}$, the actual bounds when $\eta_D$ decreases, read as
\begin{equation}
\begin{aligned}
 &  a^U_{2,\eta_D\downarrow}\approx \frac{a^U_2}{(1-\delta)^2},  \\
 &  a^L_{2,\eta_D\downarrow}\approx a^L_2(1-\delta),  \\
 &  a^U_{1,\eta_D\downarrow}\approx \frac{a^U_1}{1-\delta}-a^L_{2,\eta_D\downarrow}\cdot 2\delta,  \\
 &  a^L_{1,\eta_D\downarrow}\approx \frac{a^L_1}{1-\delta}-a^U_{2,\eta_D\downarrow}\cdot\frac{2\delta-\delta^2}{(1-\delta)^2},  \\
 &  a^U_{0,\eta_D\downarrow}\approx a^U_0-a^L_{1,\eta_D\downarrow}\cdot\delta-a^L_{2,\eta_D\downarrow}\cdot\delta^2,  \\
 &  a^L_{0,\eta_D\downarrow}\approx a^L_0-a^U_{1,\eta_D\downarrow}\cdot\delta-a^U_{2,\eta_D\downarrow}\cdot\frac{\delta^2}{1-\delta}.
\end{aligned}
\end{equation}

In conclusion, the bounds of $\{a_{n,\eta_D}\}$ under the variation of detection efficiency $\eta_D$ are given by (the detailed derivation of $\{a^U_{n,\eta_D}, a^L_{n,\eta_D}\}$ is shown in Appendix B)
\begin{equation}
\begin{aligned}
  &  a^U_{n,\eta_D} = \max\{a^U_{n,\eta_D\uparrow}, a^U_{n,\eta_D\downarrow}\},  \\
  &  a^L_{n,\eta_D} = \min\{a^L_{n,\eta_D\uparrow}, a^L_{n,\eta_D\downarrow}\},  \\
  &  (n=0,1,2).
\end{aligned}
\label{eq3}
\end{equation}

\begin{figure}[htb]
  \centering
  \includegraphics[width=8cm]{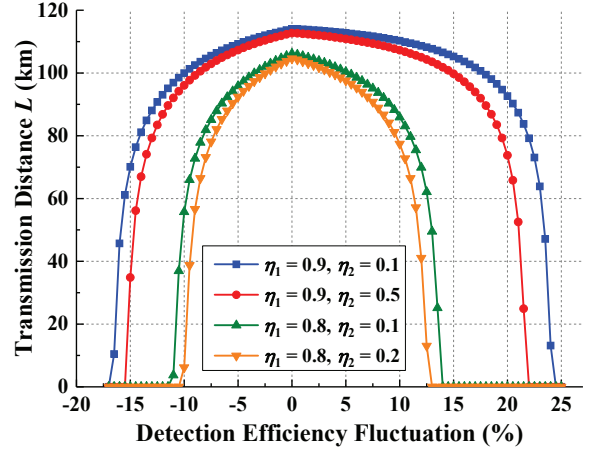}
  \caption{Transmission distance $L$ of a QKD system when the monitoring SPD's detection efficiency $\eta_D$ drifts. The original efficiency $\eta_D=10\%$ at room temperature.}\label{2-Eff}
\end{figure}

Using Eq. \ref{eq3}, we simulate $R$ and $L$ when the detection efficiency drifts, as in Fig. \ref{2-Eff}.

The original efficiency $\eta_D$ is 10\% and for either cases when it increases or decreases, the maximum transmission distance $L$ will decline. As for $\{\eta_1=0.9, \eta_2=0.1\}$, $L$ declines by 5 km when $\eta_D$ increases by over $11.5\%$ or decreases by over $5\%$, and it declines by 20 km when $\eta_D$ increases by over $19\%$ or decreases by over $12\%$. As for the worst case $\{\eta_1=0.8, \eta_2=0.2\}$, $L$ declines by 5 km when $\eta_D$ increases by over $4\%$ or decreases by over $2.5\%$, and declines by 20 km when $\eta_D$ increases by over $9\%$ or decreases by over $7\%$. The detection efficiency fluctuation here is the relative fluctuation of original detection efficiency, which is 10\% in both simulation and experiment.

\section{Experiment Results}\label{Sec3}

\begin{figure}[htb]
  \centering
  \includegraphics[width=8cm]{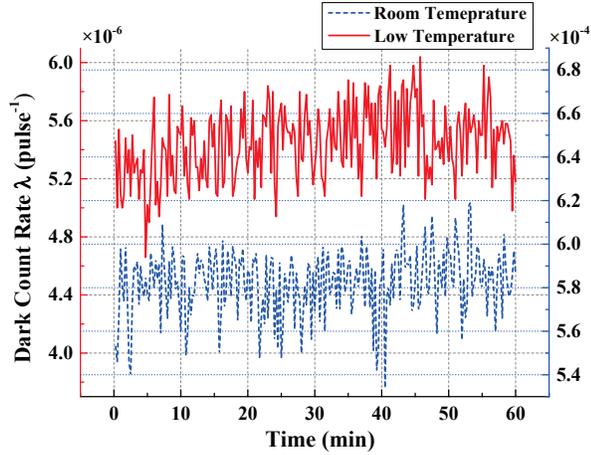}
  \caption{Drifts of DCR of the detectors at low temperature and room temperature over 60 minutes.}
  \label{3-DCR-fluc}
\end{figure}

In the experiment, we use an InGaAs SPD with its thermo-electric cooler removed to make it work at room temperature. Meanwhile, we use another detector working at low temperature (below $-30^\circ$C) for comparison~\cite{footnote2}. Each detector's DCR is measured continuously over 60 minutes, and the results are shown in Fig. \ref{3-DCR-fluc}. At low temperature, the DCR of SPD is stable, drifting from $4.6\times 10^{-6}$ to $6.1\times 10^{-6}$, while the DCR of room-temperature SPD drifts from $5.3\times 10^{-4}$ to $6.2\times 10^{-4}$, over two orders of magnitude larger than the low-temperature case.

\begin{figure}[htb]
  \centering
  \includegraphics[width=8cm]{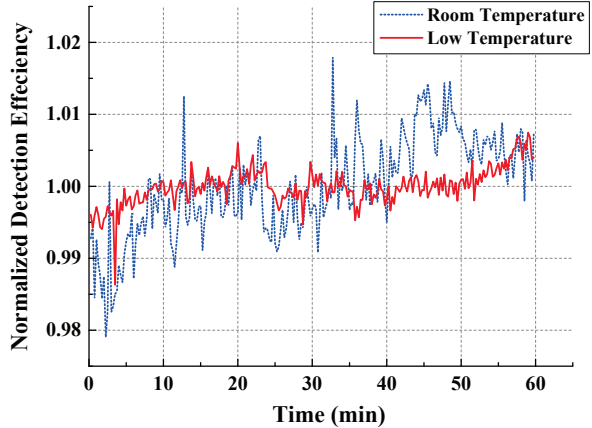}
  \caption{Drifts of detection efficiency of the detectors at low temperature and room temperature over 60 minutes.}
  \label{3-Eff-fluc}
\end{figure}

Each detector's detection efficiency is also measured over 60 minutes, and the results are shown in Fig. \ref{3-Eff-fluc}. The original detection efficiency of the two detectors are both 10\%, and the efficiency of the low temperature SPD drifts within 1\% of it in most occasions, while the efficiency drifts within 2\% of itself at room temperature.

Next, we compare the key rates when implementing PNR LSM with a low-temperature SPD and a room-temperature SPD in a practical QKD system, and we also study the key rates with modification when considering the parameter drifts, as shown in Fig. \ref{4-comparison}. According to the discussion in Appendix A, we set $\{\mu=0.6, \nu=0.1\}$ and get secret key rates under two conditions, $\{\eta_1=0.9, \eta_2=0.1\}$ and $\{\eta_1=0.8, \eta_2=0.2\}$.
\begin{figure}[htb]
  \centering
  \subfigure[~$\eta_0=1,~\eta_1=0.9,~\eta_2=0.1$]{
  \includegraphics[width=8cm]{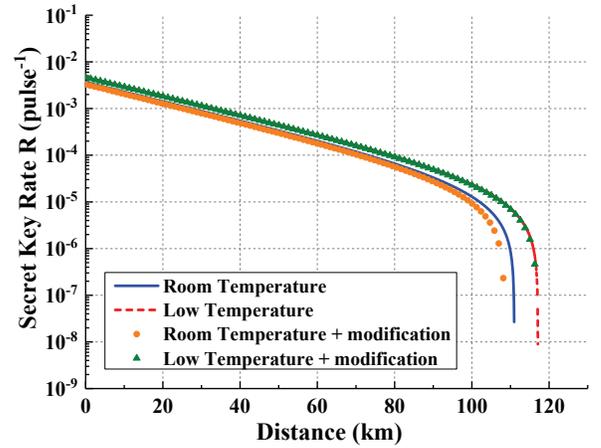}
  }
  \linebreak
  \subfigure[~$\eta_0=1,~\eta_1=0.8,~\eta_2=0.2$]{
  \includegraphics[width=8cm]{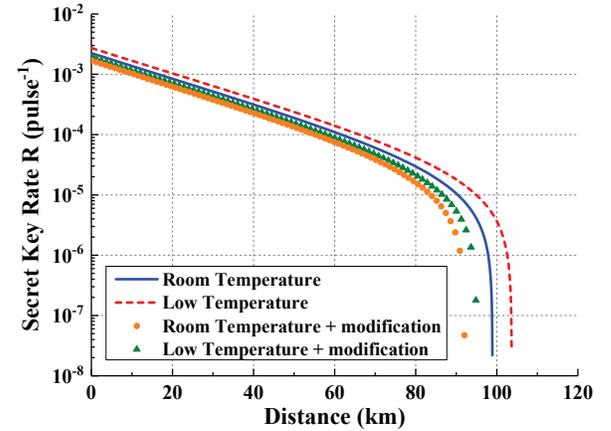}
  }
  \caption{Secret key rates of a practical QKD system under PNR LSM with a low-temperature and a room-temperature SPD in experiment, and secret key rates with modification on the parameter drifts.}
  \label{4-comparison}
\end{figure}

For the case of $\{\eta_1=0.9,\eta_2=0.1\}$, the transmission distances $L$ are 117.1 km with a low-temperature SPD and 111.0 km with a room-temperature SPD, and 116.8 km and 108.4 km respectively after being modified. As for $\{\eta_1=0.8,\eta_2=0.2\}$, the transmission distances are 103.7 km with a low-temperature SPD and 98.9 km with a room-temperature SPD, and 95.0 km and 92.0 km for the cases with modification. The secret key rates under different situations at a distance of 50 km, a typical distance in QKD networks, are also listed in Table. \ref{4-50km}.

\begin{table}[htb]\footnotesize
  \centering
  \caption{Secret key rates of the QKD system at a distance of 50 km under different situations. (Unit: per pulse)}\label{4-50km}
  \begin{tabular}{c c c}
    \hline
    \hline
     & $\mu=0.9,~\nu=0.1$ & $\mu=0.8,~\nu=0.2$ \\
    \hline
    Room Temperature & $3.35\times10^{-4}$ & $1.89\times10^{-4}$ \\
    \hline
    Room Temperature+modification & $2.97\times10^{-4}$ & $1.34\times10^{-4}$ \\
    \hline
    Low Temperature & $4.43\times10^{-4}$ & $2.37\times10^{-4}$ \\
    \hline
    Low Temperature+modification & $4.38\times10^{-4}$ & $1.54\times10^{-4}$ \\
    \hline
    \hline
  \end{tabular}
\end{table}

It can be deduced that the PNR LSM scheme has comparable monitoring performances with such two detectors for practical applications, and the difference in transmission distances is less than 7 km, which may result from the statistical fluctuations of count rates. When parameter drifts are taken into consideration, the transmission distances only decrease slightly after being modified, which are 2.6 km and 6.9 km in the two situations at room temperature, respectively. This proves that our monitoring scheme as well as the modification method are applicable in real QKD systems.

Compared with the low-temperature case, the secret key rates at room temperature have a noticeable decline almost reaching 25\% before being modified. However, the modification itself varies with $\eta_1$ and $\eta_2$ greatly. For instance, at room temperature, $R$ decreases by less than 12\% when $\{\eta_1=0.9,\eta_2=0.1\}$, and decreases by nearly 30\% when $\{\eta_1=0.8,\eta_2=0.2\}$. So values of $\eta_1$ and $\eta_2$ which insure long transmission distances and high secret key rates also insure better performances under parameter drifts of the SPD, which can be used as the criteria for choosing appropriate values of $\eta_1$ and $\eta_2$. Still, at 50 km, the secret key rate can be higher than 100 kbps in a 1 GHz QKD system with a monitoring SPD at room temperature, totally comparable with current commercial systems.

\section{Conclusion}

In summary, we propose and realize PNR LSM method by using an SPD working at room temperature, and prove that when the dark count rate of SPD is known and stable, the high DCR has no influence on monitoring performance. This can reduce the requirements of the SPD's working conditions, making it much easier to integrate the monitor module in practical QKD systems. Furthermore, the influences from the drifts of the DCR and detection efficiency of SPD are analyzed, together with a discussion of the related modification methods. We also measure such two kind of drifts at both low temperature and room temperature in real detectors, and realize the scheme in a real QKD system. The experimental results are in good agreement with the theoretical predictions and thus verify favorable prospect of using our scheme in a practical system with the analysis of parameter drifts in real circumstances.

\section{Acknowledgement}
We acknowledge Heping Zeng (ECNU) and Shengxiang Zhang (ROI Optoelectronics Tech.) for their supports on experiment. This work is supported by  the State Key Project of National Natural Science Foundation of China (Grant No. 61531003), the National Science Fund for Distinguished Young Scholars of China (Grant No. 61225003) and the Foundation of Science and Technology on Security Communication Laboratory (Grant No. 9140C110101150C11048).

~

\begin{appendix}

\section{Passive PNR LSM}

In PNR LSM, Alice chooses different $\eta\in\{\eta_0,~\eta_1,~\eta_2\}~(\eta_0>\eta_1>\eta_2)$ randomly, so $\{a^U_n,~a^L_n\}~(n=0,1,2)$ can be calculated with $P(\eta)$ of the PNR detector under different situations. When $\eta_0=1$,
\begin{equation}
  P(\eta_0=1)=(1-\lambda)a_0=(1-\lambda)P_0(\eta_0).
\end{equation}
In this situation, the upper bound $a^U_0$ and lower bound $a^L_0$ of $a_0$ are equal,
\begin{equation}
  a_0=\frac{P(\eta_0)}{1-\lambda}=P_0(\eta_0)=a^U_0=a^L_0.
\end{equation}
\begin{widetext}
And it can be derived that
\begin{eqnarray}
&&   \frac{P(\eta_1)}{1-\lambda}\geq a_0+(1-\eta_1)a_1, \\   %\tag{1.1}
&&   \frac{P(\eta_1)}{1-\lambda}\leq a_0+(1-\eta_1)a_1+(1-\eta_1)^2(1-a_0-a_1),
\end{eqnarray}
so $a^U_1$ and $a^L_1$, the upper and lower bound of $a_1$, read
\begin{eqnarray}
&&   a_1\leq \frac{P(\eta_1)-P(\eta_0)}{(1-\lambda)(1-\eta_1)}=\frac{P_0(\eta_1)-P_0(\eta_0)}{1-\eta_1}=a^U_1,\\
&&   a_1\geq \frac{P(\eta_1)-P(\eta_0)[1-(1-\eta_1)^2]-(1-\lambda)(1-\eta_1)^2}{(1-\lambda)[(1-\eta_1)-(1-\eta_1)^2]}
=\frac{P_0(\eta_1)-P_0(\eta_0)[1-(1-\eta_1)^2]-(1-\eta_1)^2}{(1-\eta_1)-(1-\eta_1)^2}=a^L_1.
\end{eqnarray}
Similarly,
\begin{eqnarray}
%\begin{split}
&&   \frac{P(\eta_2)}{1-\lambda}\geq a_0+(1-\eta_2)a_1+(1-\eta_2)^2a_2\geq a_0+(1-\eta_2)a^L_1+(1-\eta_2)^2a_2,  \\
&&   \frac{P(\eta_2)}{1-\lambda}\leq a_0+(1-\eta_2)a_1+(1-\eta_2)^2a_2+(1-\eta_2)^3(a_3+a_4+a_5+\cdots) \nonumber \\
&&   \leq[1-(1-\eta_2)^3]a_0+[1-\eta_2-(1-\eta_2)^3]a^U_1+[(1-\eta_2)^2-(1-\eta_2)^3]a_2+(1-\eta_2)^3,
\end{eqnarray}
so $a^U_2$ and $a^L_2$ read
\begin{eqnarray}
&&   a_2\leq \frac{P(\eta_2)-P(\eta_0)-(1-\lambda)(1-\eta_2)a^L_1}{(1-\lambda)(1-\eta_2)^2}=\frac{P_0(\eta_2)-P_0(\eta_0)-(1-\eta_2)a^L_1}{(1-\eta_2)^2}=a^U_2, \\
&&   a_2\geq \frac{P(\eta_2)-P(\eta_0)[1-(1-\eta_2)^3]-(1-\lambda)[1-\eta_2-(1-\eta_2)^3]a^U_1-(1-\lambda)(1-\eta_2)^3}
{(1-\lambda)[(1-\eta_2)^2-(1-\eta_2)^3]} \nonumber \\
&& = \frac{P_0(\eta_2)-P_0(\eta_0)[1-(1-\eta_2)^3]-[1-\eta_2-(1-\eta_2)^3]a^U_1-(1-\eta_2)^3}{(1-\eta_2)^2-(1-\eta_2)^3} =a^L_2.
\end{eqnarray}
\end{widetext}
$\{a'^U_n, a'^L_n\}$ can be calculated from the similar equations in decoy state~\cite{xu2010passive}.

\begin{figure}[htb]
  \centering
  \includegraphics[width=8.5cm]{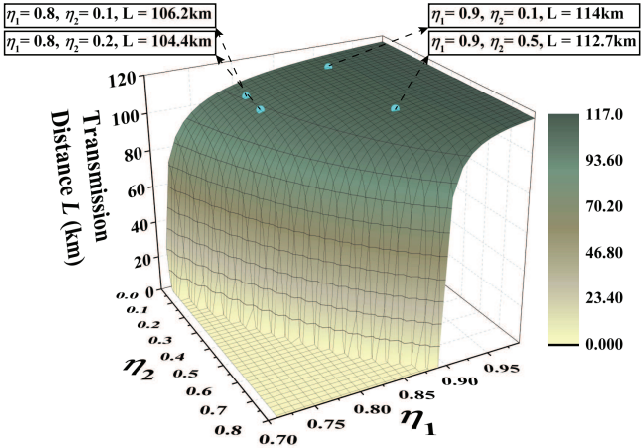}
  \caption{Transmission distance $L$ of QKD systems with PNR LSM when choosing different $\eta_1$ and $\eta_2$ ($\eta_0=1$).}\label{A-PNR}
\end{figure}

We set $\eta_2<\eta_1<\eta_0=1$, and the choosing of different $\eta_1$ and $\eta_2$ has great influence on the performance of this monitoring scheme. Fig. \ref{A-PNR} shows the maximum transmission distance $L$ of a QKD system corresponding to different $\eta_1$ and $\eta_2$. The closer $\eta_1$ is to 1 and $\eta_2$ is to 0, the longer $L$ this system has. In this paper, we set $\eta_1$ and $\eta_2$ from these typical pairs of values according to Fig. \ref{A-PNR}:
\begin{eqnarray}
&&  \eta_1=0.9, \eta_2=0.1; \eta_1=0.9, \eta_2=0.5; \nonumber \\
&&  \eta_1=0.8, \eta_2=0.1; \eta_1=0.8, \eta_2=0.2. \nonumber
\end{eqnarray}
We can find that when $\eta_0=1$, the closer $\eta_1$ is to 1 and $\eta_2$ to 0, the better the performance of PNR LSM scheme is.

\section{Calculation on the Fluctuation of DCR and Detection Efficiency}

When the DCR $\lambda$ changes into $\lambda'$, let $\lambda'=(1+\delta\lambda, |\delta\lambda|\ll 1$. As for $a^U_{0,DCR}$ and $a^L_{0,DCR}$,
\begin{equation}
  a^U_{0,DCR}=a^L_{0,DCR}=\frac{P(\eta_0)}{1-\lambda'}=a^U_0\frac{1-\lambda}{1-\lambda'}\approx a^U_0(1+\delta\lambda).
\end{equation}
And for $a^U_{1,DCR}$ and $a^L_{1,DCR}$,
\begin{equation}
\begin{aligned}
&a^U_{1,DCR}=\frac{P(\eta_1)-P(\eta_0)}{(1-\lambda')(1-\eta_1)}=a^U_1\frac{1-\lambda}{1-\lambda'}\approx a^U_1(1+\delta\lambda),\\
&a^L_{1,DCR}=\frac{P(\eta_1)-P(\eta_0)[1-(1-\eta_1)^2]-(1-\lambda')(1-\eta_1)^2}{(1-\lambda')[(1-\eta_1)-(1-\eta_1)^2]}\\
&=\Big( \frac{P(\eta_1)-P(\eta_0)[1-(1-\eta_1)^2]-(1-\lambda)(1-\eta_1)^2}{(1-\lambda)[(1-\eta_1)-(1-\eta_1)^2]}\\
&-\frac{\delta\lambda(1-\eta_1)^2}{(1-\lambda)[(1-\eta_1)-(1-\eta_1)^2]} \Big) \frac{1-\lambda}{1-\lambda'}\\
&\approx a^L_1 (1+\delta\lambda)-\delta\lambda\frac{1-\eta_1}{\eta_1}.
\end{aligned}
\end{equation}
By using similar approximation, we can get $a^U_{2,DCR}$ and $a^L_{2,DCR}$ as
\begin{equation}
\begin{aligned}
&  a^U_{2,DCR}\approx a^U_2 (1+\delta\lambda),\\
&  a^L_{2,DCR}\approx a^L_2 (1+\delta\lambda)-\delta\lambda\frac{1-\eta_2}{\eta_2}.
\end{aligned}
\end{equation}

When the detector's efficiency $\eta_D$ increases to $\eta'_D$, $(1-\eta_{BS})\eta_D>\eta_{BS}$. Let $\eta'_D = (1+\delta) \eta_D$, $|\delta\eta_D| \ll 1$, we have
\begin{equation}
  a_{n,\eta_D\uparrow} \approx \sum_{m=n}^{\infty}P_m C^n_m (1-\delta)^n \delta^{m-n}.
\end{equation}
As a result,
\begin{equation}
  a_{0,\eta_D\uparrow} = \sum_{m=0}^{\infty} a_m \delta^m = a_0 + a_1 \delta + a_2 \delta^2 + \ldots,
\end{equation}
so
\begin{equation}
\begin{split}
& a_0+a_1\delta+a_2\delta^2 < a_{0,\eta_D\uparrow}\\
& < a_0+a_1\delta+a_2(\delta^2+\delta^3+\ldots) \\
& = a_0+a_1\delta+a_2\cdot\frac{\delta^2}{1-\delta}.
\end{split}
\end{equation}
And
\begin{equation}
  a_{1,\eta_D\uparrow} = \sum_{m=1}^{\infty} a_m(1-\delta)\delta^{m-1},
\end{equation}
so
\begin{equation}
\begin{split}
& (1-\delta)(a_1+a_2\cdot 2\delta)<a_{1,\eta_D\uparrow}\\
& < (1-\delta)[a_1+a_2(\sum_{m=2}^{\infty}m\delta^{m-1})]\\
& = (1-\delta)[a_1+a_2\cdot \frac{2\delta-\delta^2}{(1-\delta)^2}].
\end{split}
\end{equation}
Similarly,
\begin{equation}
  a_{2,\eta_D\uparrow} = \sum_{m=2}^{\infty} a_m\frac{m(m-1)}{2}(1-\delta)^2\delta^{m-2},
\end{equation}
so
\begin{equation}
\begin{split}
& (1-\delta)^2a_2<a_{2,\eta_D\uparrow}\\
& < (1-\delta)^2[a_2(\sum_{m=2}^{\infty}\frac{m(m-1)}{2}\delta^{m-2})] = \frac{a_2}{1-\delta}.
\end{split}
\end{equation}

Thus, when the detection efficiency $\eta_D$ increases, the upper bounds $\{a^U_{n,\eta_D\uparrow}\}$ and lower bounds $\{a^L_{n,\eta_D\uparrow}\}$ of the count rates of pulses with different photon numbers read ($n=0, 1, 2$)
\begin{equation}
\begin{aligned}
 &  a^U_{0,\eta_D\uparrow}=a^U_0+a^U_1\delta+a^U_2\cdot\frac{\delta^2}{1-\delta},  \\
 &  a^L_{0,\eta_D\uparrow}=a^L_0+a^L_1\delta+a^L_2\cdot\delta^2,  \\
 &  a^U_{1,\eta_D\uparrow}=(1-\delta)[a^U_1+a^U_2\cdot\frac{2\delta-\delta^2}{(1-\delta)^2}],  \\
 &  a^L_{1,\eta_D\uparrow}=(1-\delta)[a^L_1+a^L_2\cdot2\delta],  \\
 &  a^U_{2,\eta_D\uparrow}=\frac{a^U_2}{1-\delta},  \\
 &  a^L_{2,\eta_D\uparrow}=a^L_2\cdot(1-\delta)^2.
\end{aligned}
\end{equation}

When $\eta_D$ decreases to $\eta'_D$, $(1-\eta_{BS})\eta_D<\eta_{BS}$. Let $\eta'_D=(1-\sigma)\eta_D$, this is equivalent to $\eta_D=(1+\delta)\eta'_D$, where $1+\delta=\frac{1}{1-\sigma}, |\delta\eta'_D| \ll 1$. Then
\begin{equation}
  a_n \approx \sum_{m=n}^{\infty}a_{m,\eta_D\downarrow} C^n_m (1-\delta)^n \delta^{m-n}.
\end{equation}
As for $a_0$, $a_1$ and $a_2$, we have
\begin{equation}
\begin{split}
& a_{0,\eta_D\downarrow}+a_{1,\eta_D\downarrow}\delta+a_{2,\eta_D\downarrow}\delta^2<a_0\\
& <a_{0,\eta_D\downarrow}+a_{1,\eta_D\downarrow}\delta+a_{2,\eta_D\downarrow}\frac{\delta^2}{1-\delta},
\end{split}
\end{equation}
\begin{equation}
\begin{split}
& (1-\delta)(a_{1,\eta_D\downarrow}+a_{2,\eta_D\downarrow}\cdot 2\delta)<a_1\\
& <(1-\delta)[a_{1,\eta_D\downarrow}+a_{2,\eta_D\downarrow}\cdot \frac{2\delta-\delta^2}{(1-\delta)^2}],
\end{split}
\end{equation}
\begin{equation}
(1-\delta)^2\cdot a_{2,\eta_D\downarrow}<a_2<\frac{a_{2,\eta_D\downarrow}}{1-\delta}.
\end{equation}

So when $\eta_D$ decreases, the upper bounds $\{a^U_{n,\eta_D\downarrow}\}$ and lower bounds $\{a^L_{n,\eta_D\downarrow}\}$ read ($n=0, 1, 2$)
\begin{equation}
\begin{aligned}
 &  a^U_{2,\eta_D\downarrow}=\frac{a^U_2}{(1-\delta)^2},  \\
 &  a^L_{2,\eta_D\downarrow}=a^L_2(1-\delta),  \\
 &  a^U_{1,\eta_D\downarrow}=\frac{a^U_1}{1-\delta}-a^L_{2,\eta_D\downarrow}\cdot 2\delta,  \\
 &  a^L_{1,\eta_D\downarrow}=\frac{a^L_1}{1-\delta}-a^U_{2,\eta_D\downarrow}\cdot\frac{2\delta-\delta^2}{(1-\delta)^2},  \\
 &  a^U_{0,\eta_D\downarrow}=a^U_0-a^L_{1,\eta_D\downarrow}\cdot\delta-a^L_{2,\eta_D\downarrow}\cdot\delta^2,  \\
 &  a^L_{0,\eta_D\downarrow}=a^L_0-a^U_{1,\eta_D\downarrow}\cdot\delta-a^U_{2,\eta_D\downarrow}\cdot\frac{\delta^2}{1-\delta}.
\end{aligned}
\end{equation}

\end{appendix}
~


\begin{thebibliography}{99}
\bibitem [40] {footnote1}  We assume the photon number distribution of the light source follows Possion distribution in simulation.
\end{thebibliography}
\end{document}